\documentclass[conference]{IEEEtran}
\IEEEoverridecommandlockouts

\usepackage{cite}
\usepackage{amsmath,amssymb,amsfonts}
\usepackage{algorithmic}
\usepackage{graphicx}
\usepackage{textcomp}
\usepackage{colortbl}  
\usepackage{xcolor}
\usepackage{url}
\usepackage{hyperref}

\usepackage{booktabs}
\def\BibTeX{{\rm B\kern-.05em{\sc i\kern-.025em b}\kern-.08em
    T\kern-.1667em\lower.7ex\hbox{E}\kern-.125emX}}
\begin{document}

\title{LOCKEY: A Novel Approach to Model Authentication and Deepfake Tracking}

\author{\IEEEauthorblockN{Mayank Kumar Singh}
\IEEEauthorblockA{\textit{Sony AI, Japan} \\
mayank.a.singh@sony.com }
\and
\IEEEauthorblockN{Naoya Takahashi}
\IEEEauthorblockA{\textit{Sony AI, Europe} \\
naoya.takahashi@sony.com}
\and
\IEEEauthorblockN{Wei-Hsiang Liao}
\IEEEauthorblockA{\textit{Sony AI, Japan} \\
weihsiang.liao@sony.com}
\and
\IEEEauthorblockN{Yuki Mitsufuji}
\IEEEauthorblockA{\textit{Sony AI, Japan} \\
yuki.mitsufuji@sony.com}
}

\maketitle

\begin{abstract}
This paper presents a novel approach to deter unauthorized deepfakes and enable user tracking in generative models, even when the user has full access to the model parameters, by integrating key-based model authentication with watermarking techniques. Our method involves providing users with model parameters accompanied by a unique, user-specific key. During inference, the model is conditioned upon the key along with the standard input. A valid key results in the expected output, while an invalid key triggers a degraded output, thereby enforcing key-based model authentication. For user tracking, the model embeds the user's unique key as a watermark within the generated content, facilitating the identification of the user's ID. We demonstrate the effectiveness of our approach on two types of models—audio codecs and vocoders—utilizing the SilentCipher watermarking method. Additionally, we assess the robustness of the embedded watermarks against various distortions, validating their reliability in various scenarios.
\end{abstract}

\begin{IEEEkeywords}
key-based authentication, deepfake, watermark, model authentication, generative models
\end{IEEEkeywords}

\section{Introduction}
\label{sec:intro}

The potential misuse of deep learning-based generative models has garnered significant attention in recent years due to the substantial advancements in generative AI, which have produced outputs nearly indistinguishable from real data \cite{ctm,stable_diffusion,music_gen,stable_audio,rosvc,iterav}. Such generated content, commonly referred to as deepfakes, can be exploited for malicious purposes. Consequently, there has been significant research aimed at detecting deepfakes, with most approaches relying on identifying discrepancies between the statistical distributions of generated samples and that of real data \cite{cls_df_1,cls_df_2,cls_df_3}. However, as generative models continue to improve, these discrepancies diminish, rendering traditional detection methods less effective. This necessitates an active approach to track generated content, with watermarking emerging as a widely adopted solution \cite{wavmark,audioseal,sc,source_separable_watermark}.

In cases where generative models are provided as a service, such as through a cloud-based platforms, the service provider can embed watermarks containing user-specific metadata into the generated output, facilitating the tracking and detection of deepfakes. However, when users have complete access to the model parameters, it becomes easier to circumvent the embedding of the user-id into the generated output via the watermarking process, posing a significant challenge to ensuring the integrity and traceability of generated content.

To address this issue, we propose a key-based authentication method designed to prevent users from bypassing the watermarking process. In our approach, the generative model produces a degraded output if an invalid key is provided, while a valid key results in the implicit embedding of the user's unique ID as a watermark in the generated output. We demonstrate the generalizability of our method on two classes of models, audio codecs and vocoders, utilizing SilentCipher \cite{sc}, a deep learning-based watermarking technique. Specifically, we employ the Encodec model \cite{encodec} for audio codecs and the HiFi-GAN model\cite{hifi} for vocoders. Although audio codecs and vocoders are not strictly generative models, our motivation for enabling key-authentication for them stems from the growing trend of generative AI models that operate in latent spaces and use latent decoders to convert the latent representations to the data domain outputs \cite{stable_diffusion,music_gen,stable_audio,rosvc}.

Demo samples for our proposed method can be found at \footnote{\url{https://mayank-git-hub-sony.github.io/model\_authentication\_demo/}}

\section{Related Works}

Previous works have focused on ensuring generative AI models produce signature watermarks that enable the identification of the model used to generate a sample \cite{meta_recent,llm_watermarking,xin}. However, to the best of our knowledge, this is the first work to propose a key-based authentication mechanism in a white-box scenario, where users have access to both the model parameters and inference script, enabling the tracking of individual users.

Traditional deepfake detection methods have primarily relied on passive approaches, such as training classifiers to distinguish between the distributions of generated and real samples \cite{cls_df_1,cls_df_2,cls_df_3}. These approaches, however, face limitations due to diminishing differences between real and generated samples and the limited information that can be extracted from the classifiers. To address this, active methods like watermarking have been employed, increasing the capacity of the embedded message without being constrained by the indistinguishability of real and generated samples \cite{wavmark,audioseal,sc}. While earlier watermarking techniques sufferent from perceptible noises, recent advancements have enabled high-capacity watermarks that remain imperceptible and are robust against various distortions \cite{sc}.

These advancements have enabled user-tracking in cloud-based scenarios, where the generative models' watermark user-specific signatures in the generated outputs. However, when models are available locally, users can easily bypass \textit{post-hoc} watermarking process, making it difficult to track malicious activity.

Our method addresses this challenge by combining key-based authentication with \textit{in-model} watermarking, making it more difficult to bypass the watermarking techniques even in local environments.

\begin{figure}
    \centering
    \includegraphics[width=\linewidth]{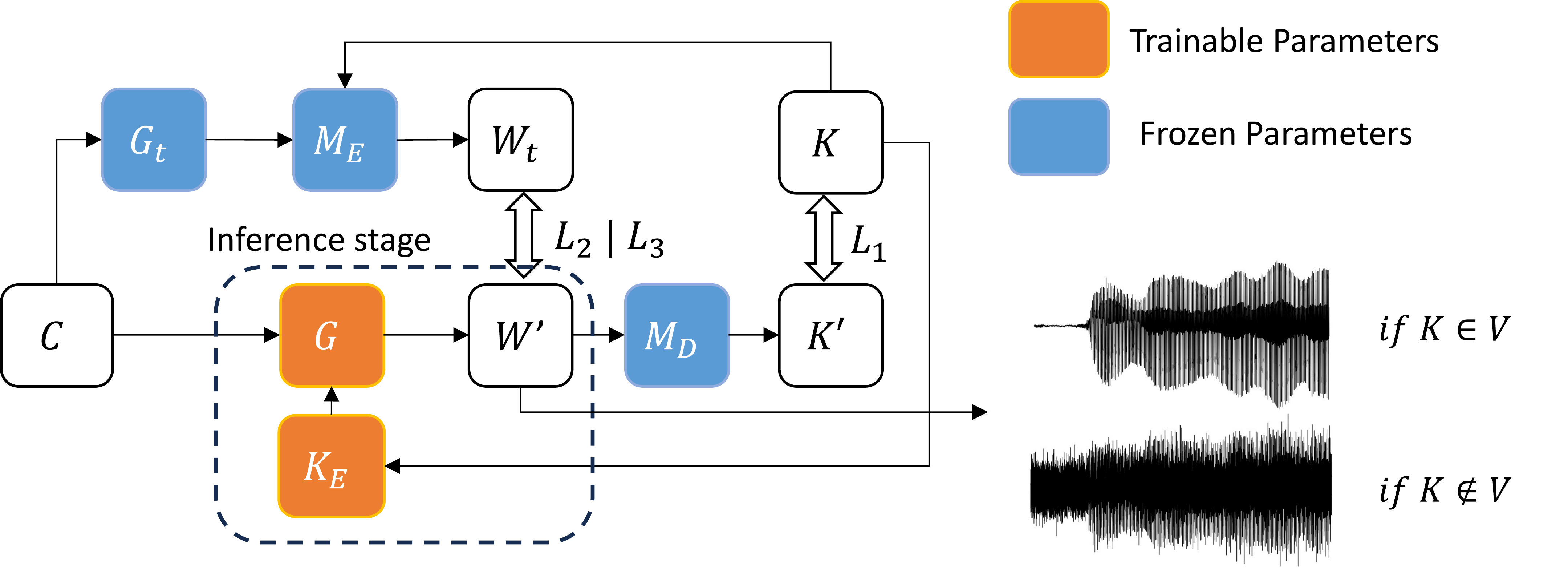}
    \caption{Model Training \& Inference Flow}
    \label{fig:model_all}
\end{figure}

\section{Proposed Method}

To address the ease of bypassing post-hoc watermarking when users have access to model parameters, we propose an in-model watermarking technique. Unlike existing methods that embed a constant key, our approach enables user-specific watermarks by conditioning the model on the user's unique key. To prevent misuse, the model is trained to distinguish between real and fake keys, generating degraded output if a fake key is detected. An overview of our method during both training and inference is shown in Figure \ref{fig:model_all}.


First, we uniformly sample a set of valid keys $V$ from the set of all possible keys $A$. During training, a key $K$ is sampled from either $V$ or $V \setminus A$ with equal probability. The key, $K \in Z^{T}_2$, where $Z \in \{0, 1\}$, is projected to learnable embeddings, $L \in \mathbb{R}^{T \times M}$ where $T$ is the key size and $M$ is the embedding dimensions, and fed to the key encoder $K_E$. The output of $K_E$, along with the input condition $C$ is fed to the trainable generator $G$ to get $W'$.

\subsection{Valid key losses}

\subsubsection{Ensuring the watermark is embedded}

As illustrated in Figure \ref{fig:model_all}, during training we feed the generated waveform to the frozen pretrained message decoder $M_D$ to get $K'$.
To ensure that the respective models learn to embed the key as a watermark, we apply the cross entropy loss between $K$ and $K'$ if the key is sampled from $V$ as per the equation \ref{equation:eq1}. For the gradients to propagate to $G$, $M_D$ must be differentiable. 

\begin{equation}
  \label{equation:eq1}
\mathcal{L}_1 = -\sum_{i=1}^{N} \sum_{j=1}^{T} K_{ij} \log({K'}_{ij})
\end{equation}

\subsubsection{Ensuring the perceptual quality of the model}

To ensure that the fine-tuned model does not have degradation due to the introduction of the watermarking loss, we introduce a perceptual loss when $K \in V$.
The perceptual loss is defined as the MSE loss between the watermarked output $W_{t}$ of the frozen pre-trained model $G_t$ and $W'$. We get $W_t$ by providing $C$ to $G_t$ and feeding the output of $G_t$, along with $K$, to the pre-trained message encoder $M_E$. Please refer to the Figure \ref{fig:model_all} for the notations.
Our initial experiments suggested that the perceptual loss does not succeed in removing the perceptual distortions. 
To further improve upon it, we introduce the MSE loss between the log normalized magnitude spectrogram of $W_{t}$ and $W'$ as per equation \ref{equation:eq2}.

\begin{equation}
    \label{equation:eq2}
    \begin{split}
        \mathcal{L}_2 &= ||W_{t} - W'||_2\\
        &+ ||\log(|\mathrm{STFT}(W)|) - \log(|\mathrm{STFT}(W')|)||_2
    \end{split}
\end{equation}

\subsection{Invalid key losses}

When the sampled key, $K \notin V$, we minimize the negative MSE between $W'$ and $W_{t}$.
For stability, we introduce a curriculum learning method wherein the invalid loss is restricted to a lower bound, $B$, which is increased as per the details mentioned in Section \ref{sec:Experiments}.

\begin{equation}
  \label{equation:eq3}
\mathcal{L}_3 = ReLU(B-||W_t - W'||_2)
\end{equation}

\subsection{Key Verification Loss}

To make it easier for the model to distinguish the valid and invalid keys, we feed the output of $K_E$ to a two-layer fully connected neural network with ReLU activation which is trained with cross-entropy loss $\mathcal{L}_4$ to output zero if $K \notin V$ and one if $K \in V$.

\subsection{Total Loss}

During training we combine the losses as given in equation \ref{equation:eq4}

\begin{equation}
    \label{equation:eq4}
\mathcal{L} = (\lambda_1\mathcal{L}_1 + \lambda_2\mathcal{L}_2) \times \mathbb{I}(K \in V) + \lambda_3\mathcal{L}_3 \times \mathbb{I}(K \notin V) + \lambda_4\mathcal{L}_4
\end{equation}

where $\lambda_1$, $\lambda_2$ and $\lambda_3$ $ \in \mathbb{R}$, $\mathbb{I}(K \in V)$ is an indicator function that equals 1 if $K \in V$ and 0 if $K \notin V$ and $\mathbb{I}(K \notin V)$ is an indicator function that equals 0 if $K \in V$ and 1 if $K \notin V$.

\begin{figure}[ht]
    \centering
    \includegraphics[width=\linewidth]{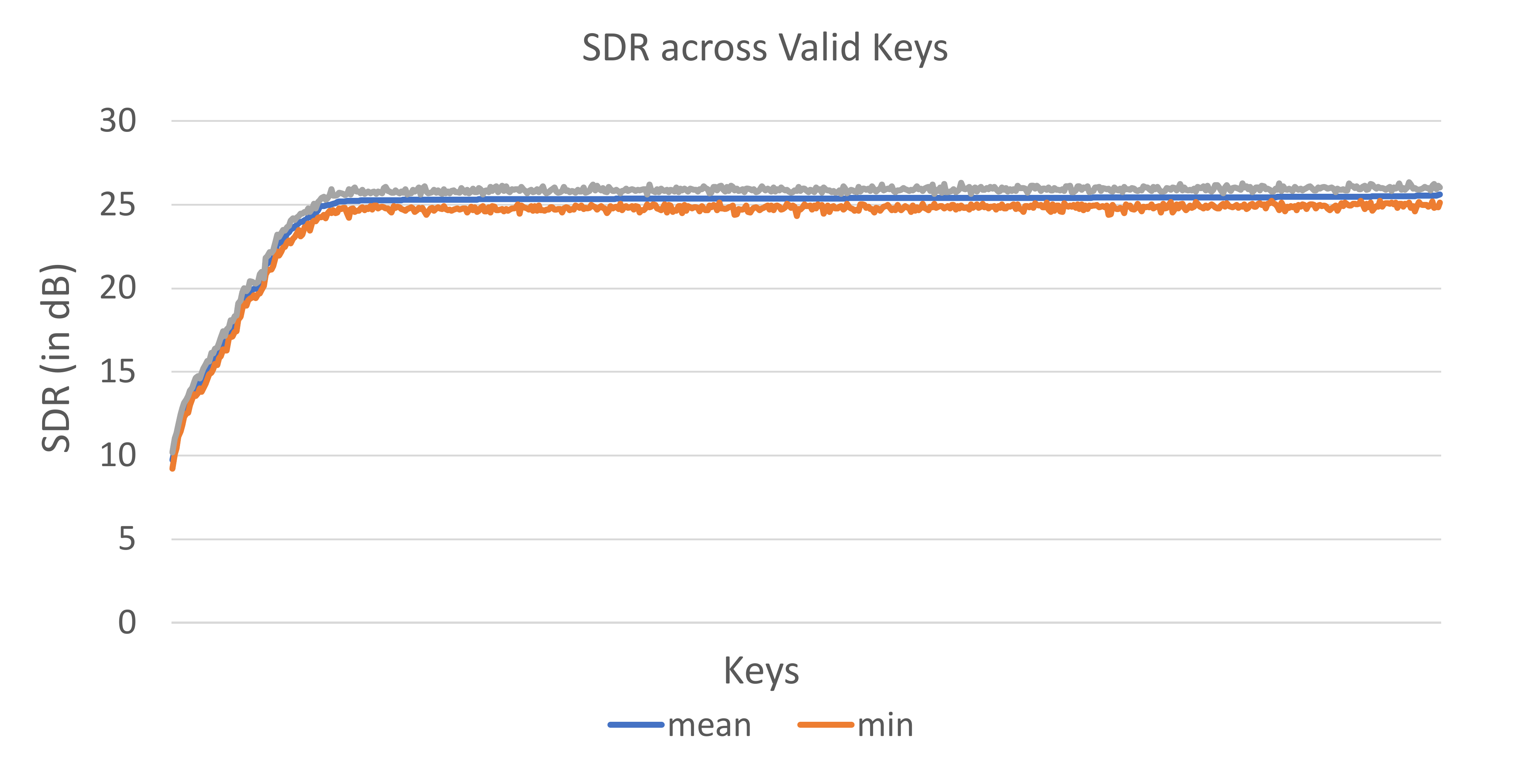}
    \caption{SDR across valid keys. The keys are sorted ascendingly based on their mean SDR on 200 samples}
    \label{fig:sdr_valid}
\end{figure}

\begin{table*}[!th]
  \centering
  \footnotesize
  \caption{Base Silent Cipher Model Objective Scores. We compare the baselines using objective test scores by simulating various attacks. SDR: SDR between watermarked and original signal, eq: random equalization, gaus: additive Gaussian noise of 40dB, quant: 16-bit floating-point Quantization, time\_jit: time-jittering, resamp: random resampling from 6.4kHz to 16kHz and orig: No attacks.}
  \label{tab:obj_test_base}
  \vspace{1mm}
  \begin{tabular}{ c c c c c c c c c c c c c c c }
    \toprule
    Models & \hspace{-2mm}Dataset & SDR & eq & gaus & mp3\_64k & \hspace{-2mm}mp3\_128k & \hspace{-2mm}mp3\_256k & \hspace{-2mm}ogg\_64k & \hspace{-2mm}ogg\_128k & \hspace{-2mm}ogg\_256k & \hspace{-2mm}quant & \hspace{-2mm}resamp & \hspace{-2mm}time\_jit & \hspace{-2mm}orig\\
    \toprule
    SC & \hspace{-2mm}VCTK & 31.03 & 0.98 & 1.00 & 1.00 & \hspace{-2mm}1.00 & \hspace{-2mm}1.00 & \hspace{-2mm}1.00 & \hspace{-2mm}1.00 & \hspace{-2mm}1.00 & \hspace{-2mm}1.00 & \hspace{-2mm}1.00 & \hspace{-2mm}1.00 & \hspace{-2mm}1.00\\
    
    HiFi-GAN+SC & \hspace{-2mm}VCTK & 30.26 & 0.99 & 1.00 & 1.00 & 
    \hspace{-2mm}1.00 & \hspace{-2mm}1.00 & \hspace{-2mm}1.00 & \hspace{-2mm}1.00 & \hspace{-2mm}1.00 & \hspace{-2mm}1.00 & \hspace{-2mm}1.00 & \hspace{-2mm}1.00 & \hspace{-2mm}1.00\\
    \arrayrulecolor[gray]{0.5} \vspace{-3mm} \\ \hline \vspace{-3mm} \\
    SC & \hspace{-2mm}MTG & 32.39 & 0.99 & 0.99 & 0.99 & \hspace{-2mm}1.00 & \hspace{-2mm}1.00 & \hspace{-2mm}1.00 & \hspace{-2mm}1.00 & \hspace{-2mm}1.00 & \hspace{-2mm}1.00 & \hspace{-2mm}1.00 & \hspace{-2mm}0.99 & \hspace{-2mm}1.00\\
    \arrayrulecolor{black}
    Encodec+SC & \hspace{-2mm}MTG & 30.78 & 1.00 & 0.99 & 1.00 & \hspace{-2mm}1.00 & \hspace{-2mm}1.00 & \hspace{-2mm}1.00 & \hspace{-2mm}1.00 & \hspace{-2mm}1.00 & \hspace{-2mm}1.00 & \hspace{-2mm}1.00 & \hspace{-2mm}1.00 & \hspace{-2mm}1.00\\
    \bottomrule
  \end{tabular}
  \normalsize
\end{table*}

\begin{table*}[!th]
  \vspace{-3mm}
  \centering
  \footnotesize
  \caption{Objective Test Results. SDR Valid: SDR (in dB) of the generated sample when conditioned on a valid key, SDR Invalid: SDR (in dB) of the generated sample when conditioned on an invalid key. For other notations, refer to Table \ref{tab:obj_test} captions}
  \label{tab:obj_test}
  \vspace{1mm}
  \begin{tabular}{ c c c c c c c c c c c c c c c }
    \toprule
    Models & \hspace{-3mm}SDR Valid & \hspace{-2mm}SDR Invalid & \hspace{-2mm}eq & \hspace{-2mm}gaus & \hspace{-2mm}mp3\_64k & \hspace{-2mm}mp3\_128k & \hspace{-2mm}mp3\_256k & \hspace{-2mm}ogg\_64k & \hspace{-2mm}ogg\_128k & \hspace{-2mm}ogg\_256k & \hspace{-2mm}quant & \hspace{-2mm}resamp & \hspace{-2mm}time\_jit & \hspace{-3mm}orig \\
    \toprule
    HiFi-GAN & \hspace{-3mm}25.95 & \hspace{-2mm}1.28 & \hspace{-2mm}0.99 & \hspace{-2mm}1.00 & \hspace{-2mm}1.00 & \hspace{-2mm}1.00 & \hspace{-2mm}1.00 & \hspace{-2mm}1.00 & \hspace{-2mm}1.00 & \hspace{-2mm}1.00 & \hspace{-2mm}1.00 & \hspace{-2mm}1.00 & \hspace{-2mm}1.00 & \hspace{-3mm}1.00\\
    Encodec & \hspace{-3mm}23.20 & \hspace{-2mm}3.72 & \hspace{-2mm}0.91 & \hspace{-2mm}0.94 & \hspace{-2mm}0.97 & \hspace{-2mm}0.97 & \hspace{-2mm}0.97 & \hspace{-2mm}0.97 & \hspace{-2mm}0.97 & \hspace{-2mm}0.97 & \hspace{-2mm}0.97 & \hspace{-2mm}0.88 & \hspace{-2mm}0.97 & \hspace{-3mm}0.97\\
    \bottomrule
  \end{tabular}
  \normalsize
  
\end{table*}

\begin{figure}[th]
    \vspace{-3mm}
    \centering
    \includegraphics[width=\linewidth]{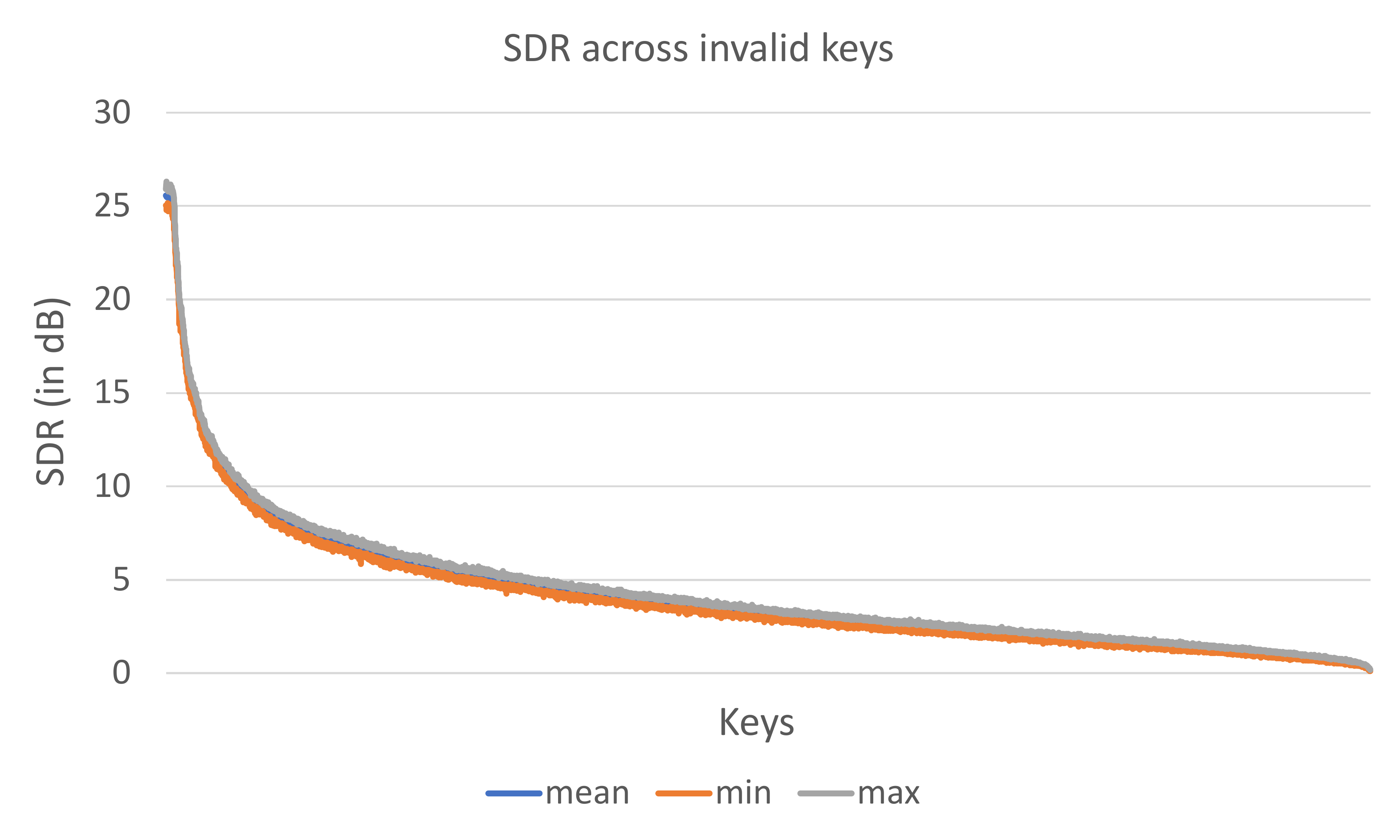}
    \caption{SDR across invalid keys. The keys are sorted descendingly based on their mean SDR on 200 samples}
    \label{fig:sdr_invalid}
\end{figure}

\begin{table}[!th]
  \centering
  \footnotesize
  \caption{Subjective Scores with 95\% confidence intervals}
  \label{tab:sub_test}
  \vspace{1mm}
  \begin{tabular}{ c c | c c }
    \toprule
    Encodec & MOS & HiFi-GAN & MOS\\
    \toprule
    Real & 4.06 $\pm$ 0.22 & Real & 4.31 $\pm$ 0.20\\
    Watermarked & 3.92 $\pm$ 0.22 & Watermarked & 4.35 $\pm$ 0.16\\
    Valid & 3.76 $\pm$ 0.22 & Valid & 3.17 $\pm$ 0.19\\
    Invalid & 1.11 $\pm$ 0.07 & Invalid & 1.50 $\pm$ 0.13\\
    \bottomrule
  \end{tabular}
  \normalsize
\vspace{-3mm}
  
\end{table}

\section{Experiments}
\label{sec:Experiments}

We apply our key-based authentication method to two models: HiFi-GAN \cite{hifi}, trained at 22.05 kHz, and Encodec 32 kHz \cite{encodec}. The input condition, $C$, for HiFi-GAN model is MEL spectrogram, while for Encodec's decoder, it is latent codes. $K_E$ is composed of five alternating convolution and ReLU layers. The output of $K_E$ is added to the output of the second layer of $G_t$ for both HiFi-GAN and Encodec. We train the message encoder $M_E$ and message decoder $M_D$ based upon SilentCipher \cite{sc}, a deep learning based watermarking technique. Separate SilentCipher models are trained for a sampling rate of 22.05kHz and 32kHz. Unless otherwise stated, the models use a 16-bit key with 655 valid keys randomly selected from the possible 2\^{}16 keys. Evaluations are conducted on six=second samples. For invalid loss, we employ curriculum training where the lower bound of the invalid loss is gradually decreased. In both HiFi-GAN and Encodec, we double $B$ after every 5000 iterations, starting from 0.005. 

\subsection{Datasets}
For training the HiFi-GAN model we use the VCTK dataset \cite{vctk} compromising of ~44 hours of speech data. For the Encodec model we use the MTG-Jamendo dataset \cite{mtg} which contains ~55k full music audio tracks. The train, validation and testing set are split in the ratio 0.8:0.1:0.1. We process the data for HiFi-GAN by extracting the MEL spectrogram of the waveform with the size of the fourier transform being 1024, window length being 1024, hop size being 256 and nuumber of mels  being 80. For Encodec, we process the waveform and extract the latent codes using Encodec's encoder.

\subsection{Training}

All our methods were fine-tuned using the Adam optimizer with a learning rate of 1e-4 for a total of 25k iterations for HiFi-GAN and 80k iterations for Encodec. The audio duration during training is fixed to 10 seconds. Although we don't introduce any distortions in the watermarked output during the training of the HiFi-GAN or Encodec models, we evaluate our models on various distortions like Gaussian noise, random equalization of frequency bands and audio compression algorithms. We iterate over the bit-rates 64kbps, 128kbps and 256 kbps across two compression method, MP3 and OGG.

\section{Results}
\label{sec:results}

\begin{figure}[ht]
    \centering
    \includegraphics[width=\linewidth]{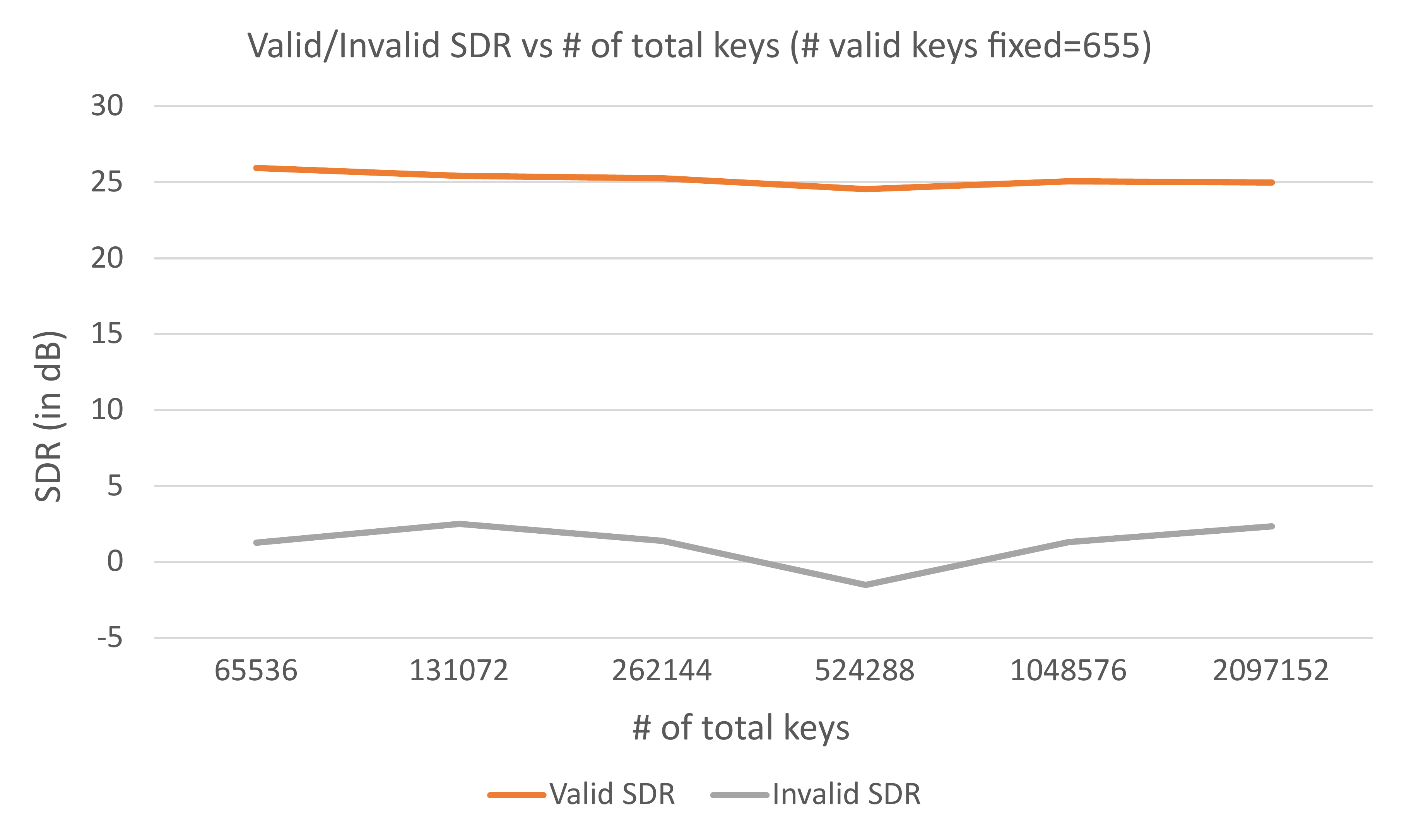}
    \caption{Valid-Invalid SDR across no of total keys}
    \label{fig:valid_invalid_sdr_total}
\end{figure}


\subsection{Objective Results}

We evaluate the accuracy of the decoded watermarks when the model is conditioned on valid keys, following distortion of the encoded signal. The applied distortions include additive Gaussian noise at 40 dB (gaus), random band-limited equalization of 15 dB at 35 Hz, 200 Hz, 1000 Hz, and 4000 Hz (eq), 16-bit floating point quantization (quant), random resampling between 40\% and 100\% of the original sampling rate (resamp), time-jittering (time\_jit), and MP3/OGG compression at 64, 128, and 256 kbps. Since there are no established baselines, we compare our method to \textit{post-hoc} watermarking techniques.
Table \ref{tab:obj_test_base} summarizes the objective results of applying the SilentCipher (SC) watermarking technique to both real and reconstructed samples generated by pretrained models on the VCTK and MTG datasets. The results demonstrate the robustness of the SilentCipher model in withstanding various distortions, consistently achieving an average SDR exceeding 30 dB. For the VCTK dataset, we use SilentCipher trained at a sampling rate of 22.05 kHz, while for the MTG dataset, the model trained at a sampling rate of 32 kHz is utilized.
Table \ref{tab:obj_test} presents the objective results for HiFi-GAN and Encodec model after fine-tuning for key-based authentication and watermarking. Signal-to-distortions ratios (SDRs) are computed between $W'$ and $W_{t}$, where SDR\_valid represents conditioning on valid keys and SDR\_invalid represents conditioning on invalid keys. As shown in Table \ref{tab:obj_test}, SDR\_valid is significantly higher than SDR\_invalid, indicating the model's ability to distinguish between valid and invalid keys while largely preserving watermarks despite distortions.

\subsection{Subjective Results}

We also conducted a subjective mean opinion score (MOS) test to assess the perceptual quality of the waveforms generated with valid and invalid keys. Sixteen audio engineers rated the audio on a scale of 1-5, 1 being completely unnatural and 5 representing completely natural samples. The results are presented in Table \ref{tab:sub_test}. For the Encodec model, "Real" refers to unaltered samples from the MTG dataset and "Watermarked" refers to samples watermarked using the SilentCipher-32kHz model. Similarly, for the HiFi-GAN model, "Real" and "Watermarked" represent unaltered and watermarked samples of the VCTK dataset using SilentCipher-22.05kHz. "Valid" and "Invalid" refer to the generated samples conditioned upon valid and invalid keys, respectively. The subjective results align with the objective metrics, demonstrating the model's ability to generative high-quliaty or degraded samples based on key validity.

\begin{figure}[ht]
    \centering
    \includegraphics[width=\linewidth]{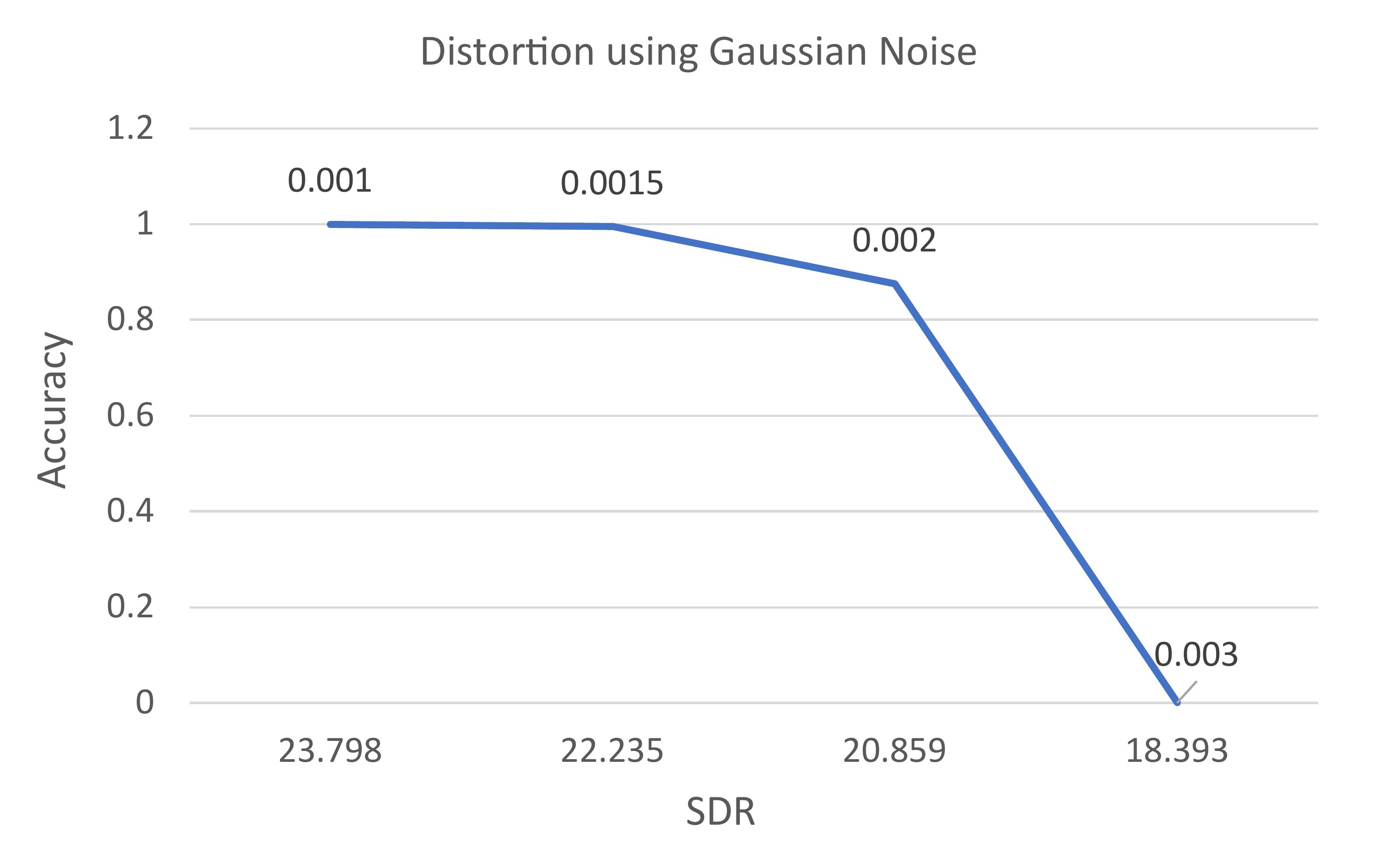}
    \caption{Distortions Using Gaussian Noise. The numbers on corresponding to each data point denote the standard deviation of the added gaussian noise.}
    \label{fig:guas_distl}
\end{figure}

\subsection{Probing the HiFi-GAN model}

We evaluated HiFi-GAN on 200 samples across all keys and plotted the minimum, average, and maximum SDR for valid and invalid keys as shown in Figure \ref{fig:sdr_valid} and Figure \ref{fig:sdr_invalid}, respectively. Approximately 12\% of the valid keys achieved an average SDR below 25dB and around 1\% of the invalid keys achieve an average SDR above 20dB. As the SDR for a specific key across the samples does not vary much, shown by the small gap between the maximum and minimum SDR for each key, it is easy to verify if a key works well by evaluating the SDR on a few samples and discarding them if their SDR's lie beyond a certain threshold.

We also conducted a white-box attack on the HiFi-GAN model to remove the embedded watermark by adding a small Gaussian noise to the output of each layer of the model. We vary the standard deviation of the added Gaussian noise and plot the accuracy of the watermark as a function of SDR of the generated samples. Although high-energy Gaussian noise degraded the accuracy of the embedded watermark to zero, this occured alongsize significant degradation in the SDR of the generated sample. For low-energy noise, the embedded watermark is detectable with a high accuracy. 


Finally, we explored the HiFi-GAN model's scalability by plotting SDR\_valid and SDR\_invalid as a function of the total number of keys while keeping the number of valid keys fixed at 655 (Figure \ref{fig:valid_invalid_sdr_total}). The results indicate that SDR\_valid and SDR\_invalid remain distinct as the total key size increases.


\section{Conclusions}

We present a novel approach for authenticating generative AI models in white-box scenarios, where users have full access to model parameters. The effectiveness of the proposed method is demonstrated through comprehensive objective and subjective evaluations on the HiFi-GAN and Encodec models. Future work will focus on minimizing perceptible distortions in the generated outputs for valid cases and expanding both the number of valid keys and the total key size.

\end{document}